\documentclass[journal]{IEEEtran}

\usepackage{amsmath}
\usepackage{graphicx}
\usepackage{mathtools}
\usepackage{amsfonts}
\usepackage{pifont}
\usepackage{amssymb}
\usepackage{epstopdf}
\usepackage{color}
\usepackage[utf8]{inputenc}
\usepackage{setspace}
\usepackage{ragged2e}
\usepackage{epsfig}
\usepackage{tabu}

\makeatletter
\newcommand{\vast}{\bBigg@{1.2}}
\newcommand{\Vast}{\bBigg@{2.3}}
\newcommand{\vastl}{\bBigg@{4}}
\newcommand{\Vastl}{\bBigg@{5}}
\newcolumntype{M}[1]{>{\centering\arraybackslash}m{#1}}
\newcolumntype{N}{@{}m{0pt}@{}}

\makeatother
\ifCLASSINFOpdf
 
\else
  
\fi

\begin{document}

\title{Security Performance Analysis of Physical Layer over Fisher-Snedecor $\mathcal{F}$ Fading Channels}

\author{Hussien Al-Hmood,~\IEEEmembership{Member,~IEEE,} 
         and H. S. Al-Raweshidy,~\IEEEmembership{Senior Member,~IEEE}
\thanks{Manuscript received May 20, 2018; xxxxx xxxxx xxxxx xxxxx xxxxx xxxxx xxxxx xxxxx xxxxx xxxxx xxxxx xxxxx xxxxx xxxxx xxxxx xxxxx xxxxx xxxxx xxxxx xxxxx xxxxx xxxxx xxxxx xxxxx xxxxx xxxxx xxxxx  xxxxx xxxxx xxxxx  xxxxx xxxxx xxxxx  xxxxx xxxxx xxxxx xxxxx xxxxx.}
\thanks{Hussien Al-Hmood is with the Department of Electrical and Electronics Engineering, University of Thi-Qar, Thi-Qar, Iraq, e-mails: hussien.al-hmood@$\{$brunel.ac.uk, eng.utq.edu.iq$\}$, h.a.al-hmood@ieee.org.}
\thanks{H. S. Al-Raweshidy is with the Department of Electronic and Computer Engineering, College of Engineering, Design and Physical Sciences, Brunel University London, UB8 3PH, U.K., e-mail: hamed.al-raweshidy@brunel.ac.uk.}}

\markboth{IEEE WIRELESS COMMUNICATIONS LETTERS,~Vol.~00, ~NO.~00, May 2018}%
{Author 1 \MakeLowercase{\textit{et al.}}: Bare Demo of IEEEtran.cls for Journals}

\maketitle

\begin{abstract}
In this letter, the performance analysis of physical layer security over Fisher-Snedecor $\mathcal{F}$ fading channels is investigated. In particular, the average secrecy capacity (ASC), the secure outage probability (SOP), the lower bound of the SOP (SOP$^L$), and the strictly positive secure capacity (SPSC) are derived in exact closed-from expressions. The Fisher-Snedecor $\mathcal{F}$ fading channel is a composite of multipath/shadowed fading that are represented by the Nakagami-$m$ distribution. Moreover, it provides close results to the practical measurements than the generalised $K$ ($K_G$) fading channels. To validate our analysis, the numerical results are affirmed by the Monte Carlo simulations.
\end{abstract}
\begin{IEEEkeywords}
Fisher-Snedecor $\mathcal{F}$ fading, average secrecy capacity, secure outage probability, strictly positive secure capacity. 
\end{IEEEkeywords}
\IEEEpeerreviewmaketitle
\section{Introduction}
\IEEEPARstart{T}{he} physical layer security of the classic Wyner's wiretap model has been widely analysed over multipath fading channels in the recent works [1]. For example, in [2] and references therein, both the main and the wiretap channels which are the Alice/Bob and Alice/Eve channels are represented by using various models of fading scenarios such as Rayleigh, Nakagami-$m$, and Rician. 
\par In a wireless communication, in addition to multipath fading, the channels may subject to the shadowing effect. Therefore, several efforts have been dedicated to study the physical layer security under composite multipath/shadowing fading scenario [2]. For instance, in [3], the average security capacity (ASC), the secure outage probability (SOP), and the strictly positive secure capacity (SPSC) over generalised-$K$ ($K_G$) fading model which is composite of Nakagami-m/Gamma distributions are derived in terms of the extended generalized bivariate Meijer G-function (EGBMGF). This is because the statistical properties, namely, the probability density function (PDF), cumulative distribution function (CDF), and the moment generating function (MGF), are derived in terms of the modified Bessel functions. Therefore, to obtain simple mathematical expressions of the performance metrics over generalised-$K$ fading channel, a mixture gamma distribution is used as an approximate framework in [4]. However, the fading parameters are assumed to be integer values. 
\par More recent, the Fisher-Snedecor $\mathcal{F}$ fading channel has been proposed as a composite of Nakagami-$m$/Nakagami-$m$ [5]. In contrast to the generalised-$K$ fading channel, the statistics of the Fisher-Snedecor $\mathcal{F}$ fading channel are derived in simple closed-form expressions. Furthermore, the Fisher-Snedecor $\mathcal{F}$ fading channel includes Nakagami-$m$, Rayleigh, and one-sided Gaussian as special cases. Therefore, it can be employed for both line-of-sight (LoS) and non-LoS (NLoS) communications scenarios with better fitting to the empirical measurements than the generalised-$K$ ($K_G$) fading model. However, it has been utilised by one work in the open technical literature [6].
\par Motivated by there is no work has been devoted to analyse the physical layer security over Fisher-Snedecor $\mathcal{F}$ fading channel, this paper investigates the aforementioned analysis. In particular, the ASC, the SOP, the lower bound of SOP (SOP$^L$), and the SPSC are derived when both the main and the wiretap channels subject to the Fisher-Snedecor $\mathcal{F}$ fading channel. To this effect and the best of the authors’ knowledge, novel analytic results of the performance metrics are obtained in exact closed-form mathematically tractable expressions. 

\section{Fisher-Snedecor $\mathcal{F}$ Fading Model}
The PDF of the received instantaneous SNR, $\gamma$, using Fisher-Snedecor $\mathcal{F}$ distribution is expressed as [5, (5)] 
\label{eqn_1}
\begin{equation}
f_{\gamma_i}(\gamma_i)=\frac{\Xi_i^{m_i}}{B(m_i,m_{s_i})}\gamma_i^{m_i-1} (1+\Xi_i\gamma_i)^{-(m_i+m_{s_i})}
\end{equation}
where $i \in \{D, E\}$, $\Xi_i=\frac{m_i}{m_{s_i} \bar{\gamma}_i}$, $m_i$, $m_{s_i}$, $\bar{\gamma}_i$ and $B(.,.)$ are the multipath index, the shape parameter, the average SNR and the beta function defined in [7, (8.380.1)], respectively.
\par The CDF of $\gamma$ using Fisher-Snedecor $\mathcal{F}$ distribution is given as [5, (4)] 
\label{eqn_2}
\begin{align}
F_{\gamma_i}(\gamma_i)&=\frac{\Xi_i^{m_i}}{m_i B(m_i,m_{s_i})}\gamma_i^{m_i} \nonumber\\
&\times{_2F_1(m_i+m_{s_i},m_i;1+m_i;-\Xi_i \gamma_i)}
\end{align}
where ${_2F_1(.,.;.;.)}$ is the hypergeometric function defined in [7, (9.14.1)].  
\section{Average Secrecy Capacity}
The ASC can be calculated by $\bar{C}_s=I_1+I_2-I_3$ [4, (6)] where $I_1$, $I_2$, and $I_3$ are given as
\label{eqn_3}
\begin{equation}
I_1=\int_0^\infty \text{ln}(1+\gamma_D)f_D(\gamma_D)F_E(\gamma_D)d\gamma_D.
\end{equation} 

\label{eqn_4}
\begin{equation}
I_2=\int_0^\infty \text{ln}(1+\gamma_E)f_E(\gamma_E)F_D(\gamma_E)d\gamma_E.
\end{equation}
\label{eqn_6_7}
\setcounter{equation}{5}
\begin{table*}
\begin{align}
I_1=&\frac{1}{\Gamma(m_D) \Gamma(m_{s_D}) \Gamma(m_E) \Gamma(m_{s_E})} \bigg(\frac{\Xi_E}{\Xi_D}\bigg)^{m_E} G^{1,1:1,2:1,2}_{1,1:2,2:2,2} 
\Bigg( \begin{matrix}
  1-m_D-m_E\\
  m_{s_D}-m_E\\
\end{matrix} \Vast\vert
\begin{matrix}
  1,1\\
  1,0\\
\end{matrix} \Vast\vert
\begin{matrix}
  1-m_E+m_{s_E}, 1-m_E\\
  0,-m_E\\
\end{matrix} \Vast\vert
\frac{1}{\Xi_D},\frac{\Xi_E}{\Xi_D}\Bigg)
\end{align}
\hrulefill
\vspace*{1pt}
\begin{align}
I_2=&\frac{1}{\Gamma(m_E) \Gamma(m_{s_E}) \Gamma(m_D) \Gamma(m_{s_D})} \bigg(\frac{\Xi_D}{\Xi_E}\bigg)^{m_D}G^{1,1:1,2:1,2}_{1,1:2,2:2,2} 
\Bigg( \begin{matrix}
  1-m_E-m_D\\
  m_{s_E}-m_D\\
\end{matrix} \Vast\vert
\begin{matrix}
  1,1\\
  1,0\\
\end{matrix} \Vast\vert
\begin{matrix}
  1-m_D+m_{s_D}, 1-m_D\\
  0,-m_D\\
\end{matrix} \Vast\vert
\frac{1}{\Xi_E},\frac{\Xi_D}{\Xi_E}\Bigg)
\end{align}
\hrulefill
\vspace*{1pt}
\end{table*}
\label{eqn_13}
\setcounter{equation}{12}
\begin{table*}
\begin{align}
\text{SOP}=&\frac{\Xi^{m_D}_D}{\Gamma(1-m_E)\Gamma(m_D) \Gamma(m_{s_D}) \Gamma(m_E) \Gamma(m_{s_E}) (1-\theta)} \bigg(\frac{\theta}{\Xi_E}\bigg)^{1+m_D} \bigg(1+\frac{\theta}{(1-\theta)\Xi_E}\bigg)^{-m_E} \bigg(1+\frac{1-\theta}{\theta} \Xi_E\bigg)^{1+m_D-m_{s_E}}\nonumber\\
& \times G^{1,1:1,1:1,2}_{1,1:1,1:2,2} 
\Bigg( \begin{matrix}
  -m_D\\
  m_{s_E}+m_E-m_D-1\\
\end{matrix} \Vast\vert
\begin{matrix}
  m_E\\
  0\\
\end{matrix} \Vast\vert
\begin{matrix}
  1-m_D+m_{s_D}, 1-m_D\\
  0,-m_D\\
\end{matrix} \Vast\vert
1+\frac{\theta}{(1-\theta)\Xi_E},\Xi_D (1-\theta)\bigg(1+\frac{\theta}{(1-\theta)\Xi_E}\bigg)\Bigg)
\end{align}
\hrulefill
\vspace*{1pt}
\end{table*}
\label{eqn_5}
\begin{equation}
\setcounter{equation}{5}
I_3=\int_0^\infty \text{ln}(1+\gamma_E)f_E(\gamma_E)d\gamma_E.
\end{equation}
\par Accordingly, $I_1$ and $I_2$ over Fisher-Snedecor $\mathcal{F}$ fading scenarios are given in (6) and (7) at the top of this page. In addition, $I_3$ is expressed as
\label{eqn_8}
\setcounter{equation}{7}
\begin{align}
I_3=&\frac{1}{\Gamma(m_E) \Gamma(m_{s_E})}G^{2,3}_{3,3} 
\Bigg( \begin{matrix}
  1-m_E,1,1\\
  m_{s_E},1,0\\
\end{matrix} \Vast\vert
\frac{1}{\Xi_E}\Bigg)
\end{align}
where $G^{a,b}_{c,d}(.)$ and $G^{a_1,b_1:...:a_n:b_n}_{c_1,d_1:...:c_n:d_n}(.)$ are Meijer G-function and EGBMGF, respectively.
\label{Proof_1}
\begin{IEEEproof}
Substituting (1) and (2) in (3), we have
\label{eqn_9}
\begin{align}
I_1&=\frac{\Xi^{m_D}_D \Xi^{m_E}_E}{m_E B(m_D,m_{s_D}) B(m_E,m_{s_E})}\nonumber\\
&\times \int_0^\infty \text{ln}(1+\gamma_D) \gamma^{m_D+m_E-1}_D (1+\Xi_D\gamma_D)^{-(m_D+m_{s_D})} \nonumber\\
&\times{_2F_1(m_E+m_{s_E},m_E;m_E+1;-\Xi_E \gamma_D)} d\gamma_D
\end{align} 
\par Invoking the identities [8, (11)], [8, (10)], and [8, (17)] with some mathematical manipulations, (9) can be rewritten as
\label{eqn_10}
\begin{align}
I_1&=\frac{\Xi^{m_D}_D \Xi^{m_E}_E}{\Gamma(m_D) \Gamma(m_{s_D}) \Gamma(m_E) \Gamma(m_{s_E})}\nonumber\\
&\times \int_0^\infty \gamma^{m_D+m_E-1}_D G^{1,2}_{2,2} \Bigg(\begin{matrix}
  1,1\\
  1,0\\
\end{matrix} \Vast\vert
 \gamma_D\Bigg)\nonumber\\
&\times G^{1,1}_{1,1} \Bigg( \begin{matrix}
  1-m_D-m_{s_D}\\
  0\\
\end{matrix} \Vast\vert
 \Xi_D\gamma_D\Bigg)\nonumber\\
&\times 
G^{1,2}_{2,2} \Bigg( \begin{matrix}
  1-m_E-m_{s_E},1-m_E\\
  0,-m_E\\
\end{matrix} \Vast\vert
 \Xi_E\gamma_D\Bigg)
 d\gamma_D
\end{align}
\par Using [9, (9)] to compute the integral in (10) and doing some mathematical simplifications, (6) is yielded which completes the proof of $I_1$.
\par Following the same steps that are employed to derive $I_1$, $I_2$ can be deduced in closed-from expression as given in (7).
\par To obtain $I_3$, we substitute (1) in (5) and recall the identity [8, (11)]. Thus, this yields
\label{eqn_11}
\begin{align}
I_3&=\frac{\Xi^{m_E}_E}{B(m_E,m_{s_E})}\nonumber\\
&\times \int_0^\infty \gamma^{m_E-1}_E (1+\Xi_E \gamma_E)^{-(m_E+m_{s_E})} G^{1,2}_{2,2} \Bigg(\begin{matrix}
  1,1\\
  1,0\\
\end{matrix} \Vast\vert
 \gamma_E\Bigg)d\gamma_E
\end{align}
\par Employing [10, (2.24.2.4)], (8) is yielded which completes the proof of $I_3$.
\end{IEEEproof}
\label{eqn_18}
\setcounter{equation}{17}
\begin{table*}
\begin{align}
\text{SOP}^L&=\frac{1}{\Gamma(m_E) \Gamma(m_{s_E}) \Gamma(m_D) \Gamma(m_{s_D})} \bigg(\frac{\theta\Xi_D}{\Xi_E}\bigg)^{m_{D}}
 G^{2,3}_{3,3} \Bigg(\begin{matrix}
  1-m_E-m_D,1-m_D-m_{s_D},1-m_D\\
  m_{s_E}-m_D,0,-m_D\\
\end{matrix} \Vast\vert
 \frac{\theta\Xi_D}{\Xi_E}\Bigg)
\end{align}
\hrulefill
\vspace*{1pt}
\end{table*} 
\section{Secure Outage Probability}
The SOP can be evaluated by [2, (14)]
\label{eqn_12}
\setcounter{equation}{11}
\begin{equation}
\text{SOP}=\int_0^\infty F_D(\theta \gamma_E+\theta-1)f_E(\gamma_E)d\gamma_E
\end{equation}
where $\theta = \mathrm{exp}(R_s) \geq 1$ with $R_s \geq 0$ is the target secrecy threshold. 
\par The SOP can be expressed in exact closed-form as given in (13) at the top of the this page.  

\label{Proof_2}
\begin{IEEEproof}
Inserting (1) and (2) in (12), the result is
\label{eqn_14}
\setcounter{equation}{13}
\begin{align}
\text{SOP}&=\frac{\Xi^{m_D}_D \Xi^{m_E}_E}{m_D B(m_D,m_{s_D}) B(m_E,m_{s_E})} \nonumber\\ 
&\times \int_0^\infty \gamma^{m_E-1}_E (\theta \gamma_E+\theta-1)^{m_D} (1+\Xi_E\gamma_E)^{-(m_E+m_{s_E})} \nonumber\\ 
&\times{_2F_1}(m_D+m_{s_D},m_D;m_D+1; \nonumber\\ 
& \hspace{3 cm} -\Xi_D (\theta\gamma_E+\theta-1)) d\gamma_E
\end{align}
\par Assuming $x=\theta \gamma_E+\theta-1$ and $dx = \theta d\gamma_E$ and performing some mathematical simplifications, (14) becomes as follows
\label{eqn_15}
\begin{align}
\text{SOP}&=\frac{\Xi^{m_D}_D }{m_D B(m_D,m_{s_D}) B(m_E,m_{s_E}) (1-\theta)}  \nonumber\\
&\times \bigg(1+\frac{\theta}{(1-\theta)\Xi_E}\bigg)^{-m_E}\bigg(1+\frac{1-\theta}{\theta}\Xi_E\bigg)^{-m_{s_E}} \nonumber\\
&\times \int_0^\infty x^{m_D} \vast(1+\frac{1}{1-\theta}x \vast)^{m_E-1}  \nonumber\\
&\times \vast(1+\frac{\Xi_E}{\theta+(1-\theta)\Xi_E} x \vast)^{-(m_E+m_{s_E})} \nonumber\\
&\times {_2F_1(m_D+m_{s_D},m_D;m_D+1;-\Xi_D x)} dx
\end{align}
\par Utilising the identities [8, (10)] and [8, (17)], (15) is expressed as 
\label{eqn_16}
\begin{align}
\text{SOP}&=\frac{\Xi^{m_D}_D }{\Gamma(1-m_E) \Gamma(m_D) \Gamma(m_{s_D}) \Gamma(m_E) \Gamma(m_{s_E}) (1-\theta)}  \nonumber\\
&\times \bigg(1+\frac{\theta}{(1-\theta)\Xi_E}\bigg)^{-m_E}\bigg(1+\frac{1-\theta}{\theta}\Xi_E\bigg)^{-m_{s_E}} \nonumber\\
&\times \int_0^\infty x^{m_D} 
G^{1,1}_{1,1} \Bigg(\begin{matrix}
  m_E\\
  0\\
\end{matrix} \Vast\vert
 \frac{1}{1-\theta}x\Bigg)\nonumber\\
&\times G^{1,1}_{1,1} \Bigg( \begin{matrix}
  1-m_E-m_{s_E}\\
  0\\
\end{matrix} \Vast\vert
 \frac{\Xi_E}{\theta+(1-\theta)\Xi_E} x\Bigg)\nonumber\\
&\times 
G^{1,2}_{2,2} \Bigg( \begin{matrix}
  1-m_D-m_{s_D},1-m_D\\
  0,-m_D\\
\end{matrix} \Vast\vert
 \Xi_D x\Bigg)dx
\end{align}
\par Making use of [9, (9)], the derived result in (13) is yielded
\end{IEEEproof}
\section{Lower Bound of the Secure Outage Probability}
The SOP$^L$ can be computed by [2, (17)]
\label{eqn_17}
\begin{equation}
\text{SOP}^L=\int_0^\infty F_D(\theta\gamma_E)f_E(\gamma_E)d\gamma_E
\end{equation}
\par The SOP$^L$ over Fisher-Snedecor $\mathcal{F}$ fading scenarios can be derived as given in (18) at the top of this page. 
\label{Proof_3}
\begin{IEEEproof}
Plugging (1) and (2) in (17) and doing some mathematical manipulations, we have 
\label{eqn_19}
\setcounter{equation}{18}
\begin{align}
\text{SOP}^L&=\frac{\theta^{m_D} \Xi^{m_D}_D}{B(m_E,m_{s_E}) \Gamma(m_D) \Gamma(m_{s_D}) \Xi^{m_{s_E}}_E}\nonumber\\
&\times \int_0^\infty \gamma^{m_E+m_D-1}_E \vast(\frac{1}{\Xi_E}+\gamma_E \vast)^{-(m_E+m_{s_E})}\nonumber\\
&\times {_2F_1(m_D+m_{s_D},m_D;m_D+1;-\theta \Xi_D x)} d\gamma_E
\end{align}
\par With the help of [8, (17)], (19) can be rewritten as 
\label{eqn_20}
\begin{align}
\text{SOP}^L&=\frac{\theta^{m_D} \Xi^{m_D}_D}{B(m_E,m_{s_E}) \Gamma(m_D) \Gamma(m_{s_D}) \Xi^{m_{s_E}}_E}\nonumber\\
&\times \int_0^\infty \gamma^{m_E+m_D-1}_E \vast(\frac{1}{\Xi_E}+\gamma_E \vast)^{-(m_E+m_{s_E})} \nonumber\\
&\times G^{1,2}_{2,2} \Bigg(\begin{matrix}
  1-m_D-m_{s_D},1-m_D\\
  0,-m_D\\
\end{matrix} \Vast\vert
 \Xi_D \theta \gamma_E\Bigg)d\gamma_E
\end{align}
\par Utilising [10, (2.23.2.4)] to compute the integral in (20), the result in (18) is deduced and this completes the proof.
\end{IEEEproof}

\section{Strictly Positive Secure Capacity}
The SPSC is expressed as [2, (20)]
\label{eqn_21}
\begin{equation}
\text{SPSC}=1-\text{SOP} \quad \text{for} \quad \theta=1 
\end{equation}
\par Consequently, the SPSC over Fisher-Snedecor $\mathcal{F}$ fading channels can be obtained by using (13) and $\theta = 1$ and inserting the result in (21).
\begin{figure}[t]
\centering
  \includegraphics[width=3.5 in, height=2.5 in]{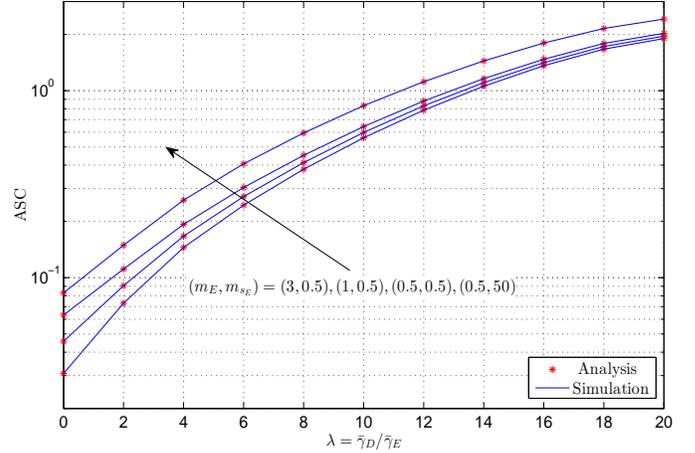}
\centering
\caption{ASC over Fisher-Snedecor $\mathcal{F}$ fading channels versus $\lambda$ for different values of ($m_E$, $m_{s_E}$), $\bar{\gamma}_E = 5$ dB, $m_D$ = 2.5, and $m_{s_D} = 5$.}
\end{figure}
\section{Analytical and Simulation Results}
In this section, to validate our derived expressions of the physical layer security over Fisher-Snedecor $\mathcal{F}$ fading channels, the Monte Carlo simulations that are obtained via generating $10^7$ realizations are compared with the analytical results. In all figures, the simulations and the numerical results of the performance metrics that are plotted versus $\lambda = \bar{\gamma}_D/\bar{\gamma}_E$ for $m_D = 2.5$ and $m_{s_D} = 5$ (moderate shadowing) are represented by the solid lines and the stars, respectively. Moreover, two different scenarios of the shadowing impact at the eavesdropper which are light and heavy shadowing are studied by using $m_{s_E} = 0.5$ and $m_{s_E} = 50$, respectively. In all results, a MATHEMATICA code that is provided in [9] has been used to calculate the EGBMGF. This is because it is not available as a built in function in MATLAB and MATHEMATICA software packages.   
\par Figs. 1-5 show the ASC, the SOP, the SOP$^L$, and the SPSC over Fisher-Snedecor $\mathcal{F}$ fading channels for $\bar{\gamma}_E = 5$ dB and different values of the fading parameters $m_E$ and $m_{s_E}$. In these figures, it can be observed that the performance becomes better, when $m_{s_E}$ increases. This is because small and large values of $m_{s_E}$ correspond to light and heavy shadowing, respectively. For instance, in Fig. 1, when $\lambda = 6$ and $m_E = 0.5$ (fixed), the ASC for $m_{s_E} = 50$ is approximately $25\%$ higher than $m_{s_E} = 0.5$. In the same context, when $m_E$ increases, the ASC decreases. This refers to less impact of the multipath on the Eve which would lead to reduce the total ASC. 
\par In Figs. 2 and 4 that are plotted for $R_s = 1$ bit/s/Hz, one can see that the values of SOP are greater than or equal to the SOP$^L$ which confirms our derived expressions. Furthermore, another confirmation that proves the validation of our analysis is the perfect matching between the numerical results and their Monte Carlo simulation counterparts in all provided figures.

\section{Conclusions}
In this letter, the secrecy performance of physical layer over Fisher-Snedecor $\mathcal{F}$ fading channels is analysed. Specifically, the ASC, the SOP, the SOP$^L$, and the SPSC are derived in exact mathematically tractable closed-form expressions. The results of this work provide a good insight about the security of the physical layer over composite multipath/shadowing fading channels when the wireless channels subject to heavy, moderate, or light shadowing. Moreover, the analysis of the physical layer security over different scenarios can be deduced from the derived expressions by setting $m$ and $m_s$ for specific values such as the Nakagami-$m$ fading condition is obtained by inserting $m_s\rightarrow \infty$ and $m = \textbf{m}$ where $\textbf{m}$ is the Nakagami-$m$ multipath index.

\begin{figure}[t]
\centering
  \includegraphics[width=3.5 in, height=2.5 in]{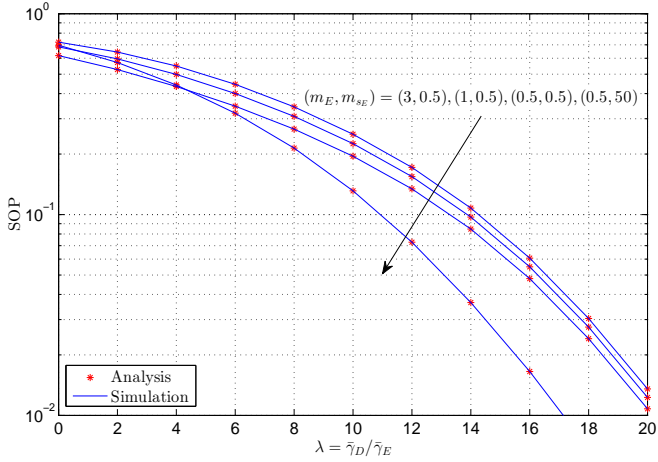}
\centering
\caption{SOP over Fisher-Snedecor $\mathcal{F}$ fading channels versus $\lambda$ for different values of ($m_E$, $m_{s_E}$), $\bar{\gamma}_E = 5$ dB, $m_D$ = 2.5, $m_{s_D} = 5$, and $R_s$ = 1.}
\end{figure} 

\begin{figure}[t]
\centering
  \includegraphics[width=3.5 in, height=2.5 in]{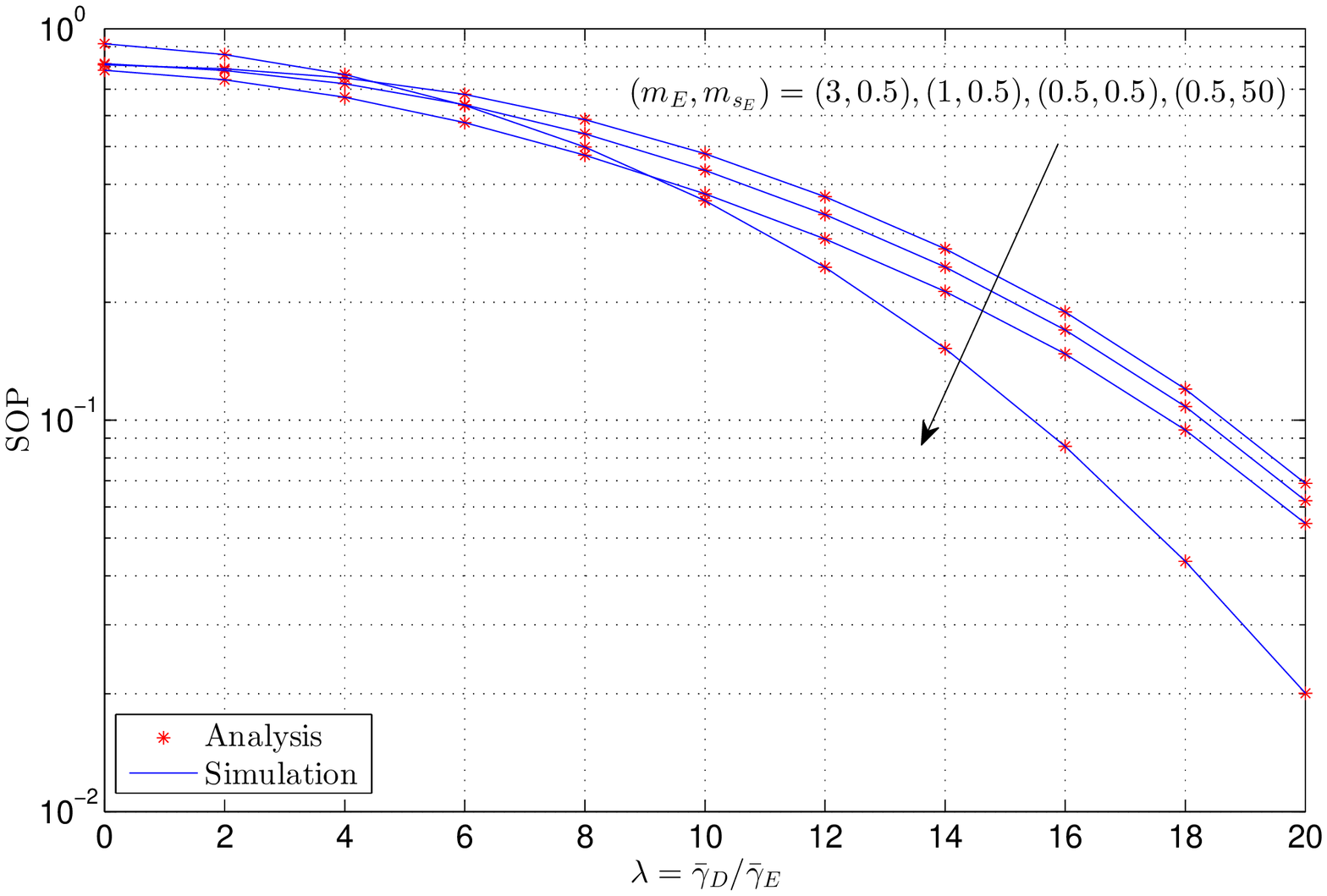}
\centering
\caption{SOP over Fisher-Snedecor $\mathcal{F}$ fading channels versus $\lambda$ for different values of ($m_E$, $m_{s_E}$), $\bar{\gamma}_E = 5$ dB, $m_D$ = 2.5, $m_{s_D} = 5$, and $R_s$ = 2.}
\end{figure} 

\begin{figure}[t]
\centering
  \includegraphics[width=3.5 in, height=2.5 in]{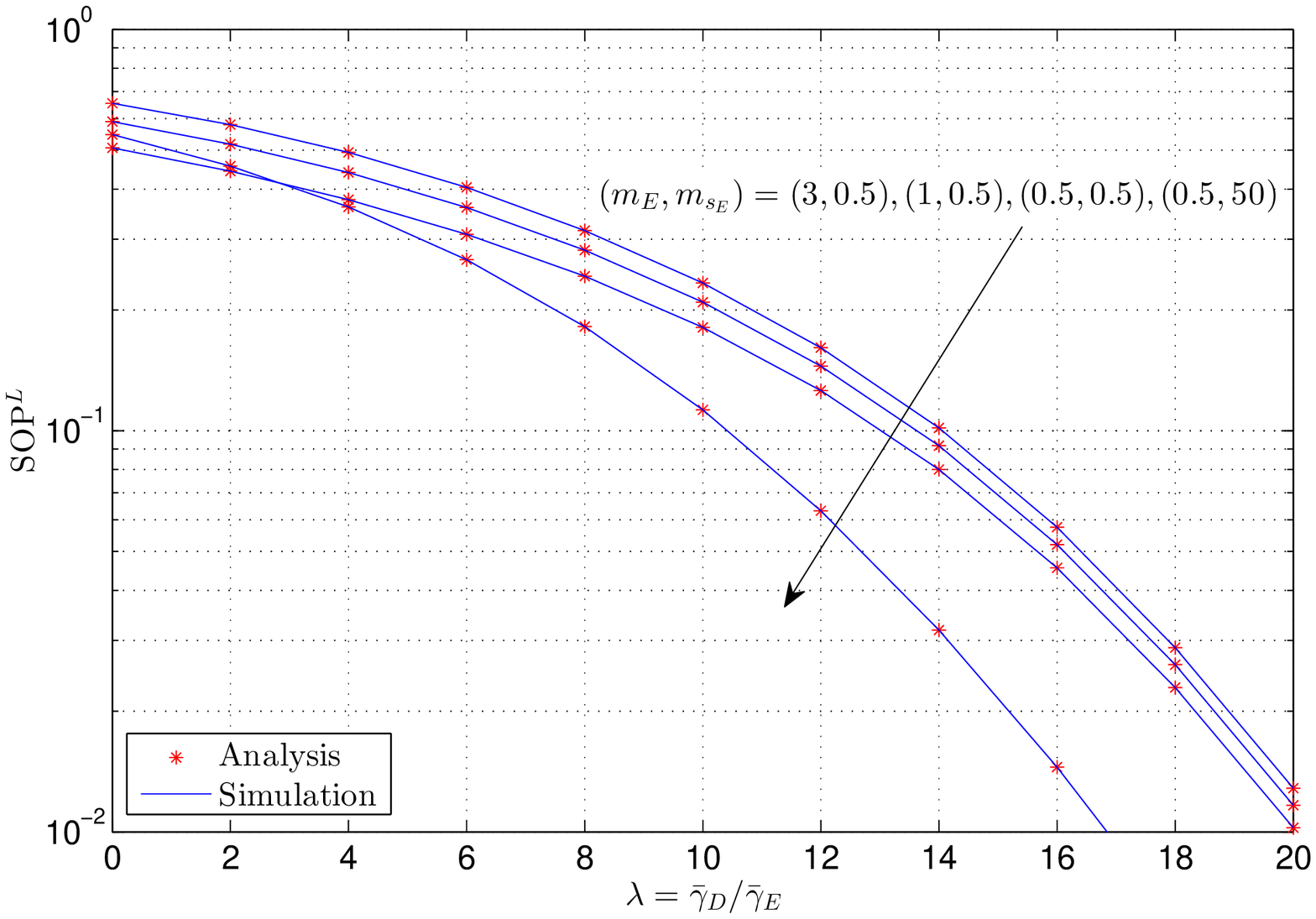} 
\centering
\caption{SOP$^L$ over Fisher-Snedecor $\mathcal{F}$ fading channels versus $\lambda$ for different values of ($m_E$, $m_{s_E}$), $\bar{\gamma}_E = 5$ dB, $m_D$ = 2.5, $m_{s_D} = 5$, and $R_s$ = 1.}
\end{figure}

\begin{figure}[t]
\centering
  \includegraphics[width=3.5 in, height=2.5 in]{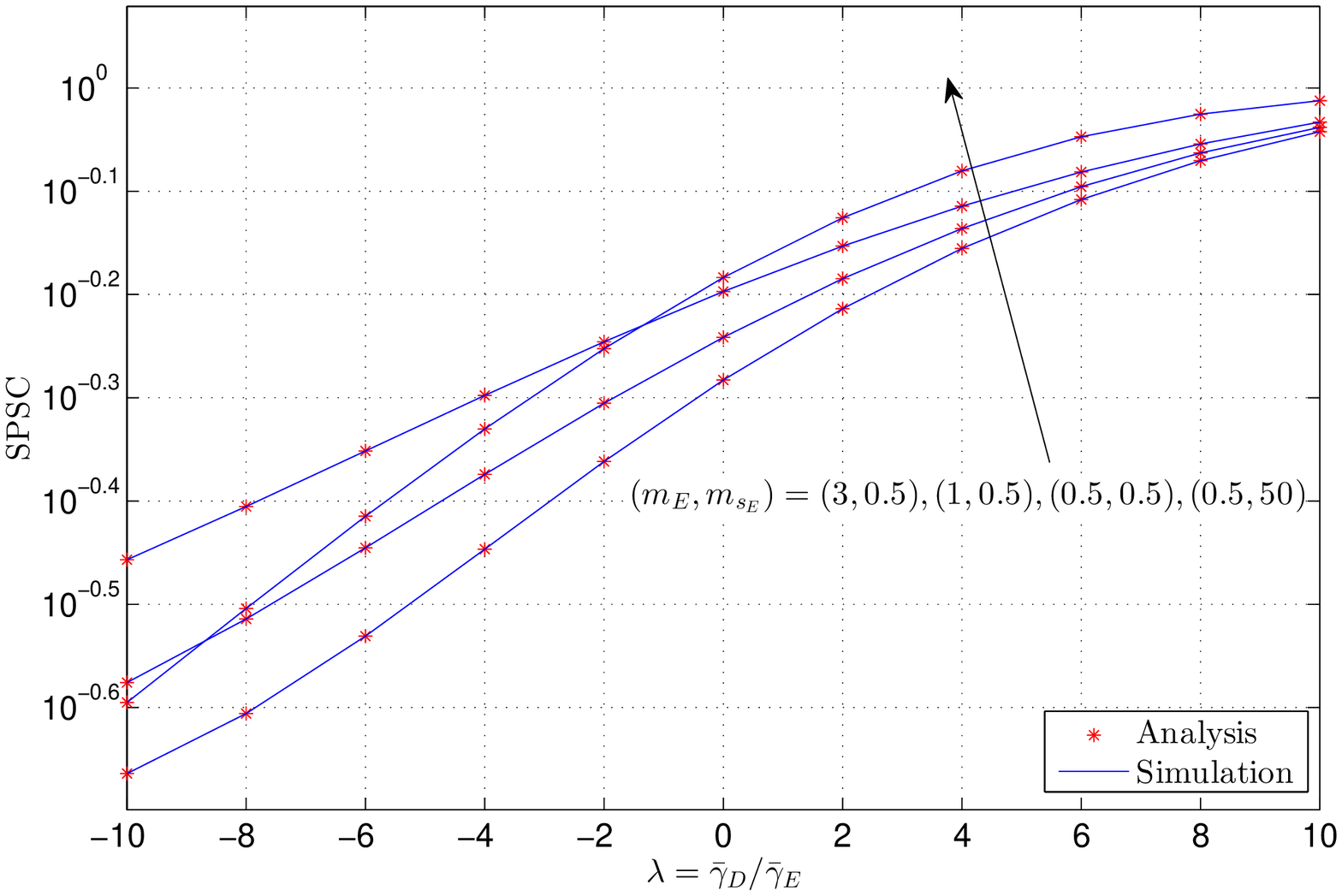} 
\centering
\caption{SPSC over Fisher-Snedecor $\mathcal{F}$ fading channels versus $\lambda$ for different values of ($m_E$, $m_{s_E}$), $\bar{\gamma}_E = 5$ dB, $m_D$ = 2.5, and $m_{s_D} = 5$.}
\end{figure}

\ifCLASSOPTIONcaptionsoff
  \newpage
\fi

\end{document}